\documentclass[aps,pra,twocolumn,footinbib,groupedaddress]{revtex4-1} 
	
\usepackage{amsmath}
\usepackage{amssymb}
\usepackage{graphicx} 
\usepackage[breaklinks=true,colorlinks,citecolor=blue,linkcolor=blue,urlcolor=blue]{hyperref}
\usepackage{physics}
\usepackage{bbold}
\usepackage{soul}
\usepackage{enumerate}
\usepackage{color}
\usepackage{ulem}
\usepackage{csquotes}
\usepackage[mathscr]{euscript}

\definecolor{dblue}{rgb}{0,0,0.6}
\definecolor{lblue}{rgb}{0.5,0.3,1}
\definecolor{lgray}{rgb}{0.6,0.6,0.6}

\begin{document}

\title{1/f noise in quantum nanoscience}

\author{G. Falci}
\affiliation{Dipartimento di Fisica e Astronomia Ettore Majorana, Universit\`a di Catania, Via S. Sofia 64, I-95123, Catania, Italy}
\affiliation{INFN, Sez. Catania, I-95123, Catania, Italy}
\affiliation{CNR-IMM, Via S. Sofia 64, I-95123, Catania, Italy}

\author{Pertti J. Hakonen}
 \affiliation{%
Low Temperature Laboratory, Department of Applied Physics, Aalto University, Puumiehenkuja 2B Otaniemi,
P.O. Box 15100, FI-00067 AALTO, Finland}
\affiliation{InstituteQ - The Finnish Quantum Institute,
Otakaari 1, FI-02150 Espoo, Finland}

\author{E. Paladino}
\affiliation{Dipartimento di Fisica e Astronomia Ettore Majorana, Universit\'a di Catania, Via S. Sofia 64, I-95123, Catania, Italy}
\affiliation{INFN, Sez. Catania, I-95123, Catania, Italy}
\affiliation{CNR-IMM, Via S. Sofia 64, I-95123, Catania, Italy}

\begin{abstract}
Fundamental issues of $1/f$ noise in quantum nanoscience are reviewed starting from basic statistical noise processes. Fundamental noise models based on two-level systems (TLS) are described.  We emphasize the importance of TLSs in materials parameter fluctuations, such as dielectric constant.  The present understanding of $1/f$ noise in superconducting quantum interferometers and in single electron devices is summarized. For coherent quantum nanoscience, we introduce superconducting qubits and the relation between decoherence and $1/f$ noise using the filter function formulation. We also clarify the qubit noise spectroscopy and emphasize the importance of materials with reduced $1/f$ noise for future quantum coherent nanodevices. 
\end{abstract}

\maketitle

{\bf Key Points} 
\begin{itemize}
\item Basic characteristics of $1/f$ noise in condensed matter systems
\item Theoretical issues concerning stationarity, ergodicity, and TLS with uniform distribution for tunneling parameter
\item Specific features of $1/f$ noise in nanoscale systems and quantum materials
\item Significance of TLS for dissipation in superconducting materials for low-loss quantum circuits
\item $1/f$ noise in quantum circuits and devices, both in the flux and charge fluctuation  regimes
\item Influence of $1/f$ noise on coherence properties of qubits
\item Role of low-frequency fluctuations in quantum sensing
\end{itemize}


\section{Introduction}
\label{introduction}

Fluctuations with power spectral density $S(f)$ increasing with decreasing frequency have been observed in a great variety of systems from physics, electronic engineering, and biology to music, economic and social sciences, even medicine. Solid-state materials and devices have formed the supreme arena where the command of the physics of low-frequency fluctuations has been deeply exploited. Strong motivation comes both from fundamental research and from the quest for electronic devices with improved performance. The advent of modern nanoscience has opened a new era in which this subject has entered the realm of quantum coherence.

Measurements of fluctuations in a vast number of physical systems 
yield a noise spectral density of the form 
\begin{equation}
    \label{eq:powerspec}
    S(f) = C \, f^{-\alpha}
\end{equation}
with $\alpha$ ranging from 0 to 2, down to the lowest measured frequencies $\sim \mu$Hz.  Since values of $\alpha \approx 1$ are typically observed, the phenomenon is called $1/f$ noise. The alternative terms "flicker noise" or "excess noise" are less frequently used. 

\begin{figure}
   \centering
  \resizebox{0.9\columnwidth}{!}{\includegraphics{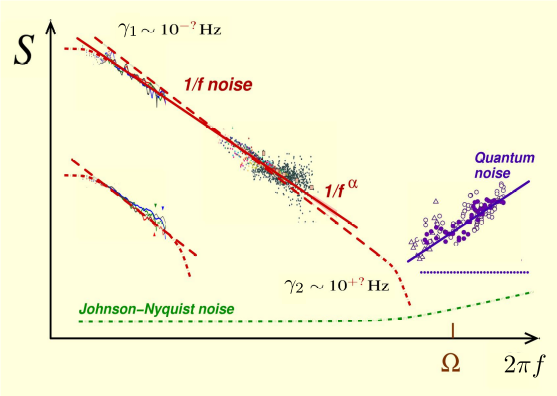}}
  \caption{
 Qualitative representation of  noise spectrum in coherent nanodevices on a log-log scale. 
 Lines describe the expected behavior in different frequency ranges. Dots represent typical observations in frequency windows depending on the noise detection method (different
 symbols and dots), see discussion in Section \ref{sec:oneoverf-coherent}. 
Long-dashed lines indicate the $1/f$ dependence between the intrinsic low- and high-frequency cutoffs $\gamma_1$ and $\gamma_2$, which in general are not known.
Thick lines indicate the $1/f^\alpha$ dependence, with $\alpha \approx 0.9$. $\Omega$ denotes the typical qubit splitting, of the order of $10^{10}$~Hz in superconducting qubits.
In this range, quantum noise is commonly white or Ohmic (dotted and thin lines). The dash-dotted line indicates the linearly increasing Johnson-Nyquist noise.
}
\label{fig:powerspec}
\end{figure}

In solid-state materials and devices fluctuations of transport and dielectric properties exhibit a $\sim 1/f$ spectrum in addition to intrinsic thermal noise and quantum (shot) noise (see Fig. \ref{fig:powerspec}). They share the same distinctive features as the frequency dependence of the noise, the time dependence of correlations, weak stationarity and other statistical properties.

With the advent of nanofabrication  technologies, the number of devices and electronic components showing excess low-frequency noise has increased, and they all share the features of the ubiquitous $1/f$ noise. Charge fluctuations with $1/f$ spectrum in substrate materials have been found to influence devices operating in the Coulomb blockade regime where quantized charge in a small, weakly-coupled conducting island is acting as the quantum mechanical observable. In the regime of superconducting coherence, flux noise generates unwanted currents with $1/f$ character, which is detrimental in many magnetic field sensors and superconducting qubits. Also, various heterostructures (including e.g. field effect (FET) and high-electron-mobility transistor (HEMT) structures), either based on materials deposition, modulation doping, or stacking of atomically layered two-dimensional (2D) materials, display $1/f$ noise in the number of charge carriers owing to charge traps in the vicinity of the active device channel. Beyond electronic and physical systems, $1/f$ noise is found even in healthy human heartbeat rates.

The appearance of $1/f$-like  behavior in many  phenomena of diverse nature and its scale invariance, i.e. the fact that each decade of frequencies contains the same integrated power, suggested that a common, deeper level  description may exist, such as for the universality of the exponents in critical phenomena, i.e. that $1/f$-like behavior is an inherent property of all systems and that a universal minimum frequency exists. However, the large amount of experimental data in solid-state systems indicates that there is no universal spectrum of low-frequency noise.

In modern electronic devices with small active volume 
bistable switching of the resistance called random telegraph noise (RTN) is observed. The phenomenon is attributed to capture and emission processes from defects suggesting a possible microscopic mechanism of $1/f$ noise. In fact, in many solid-state systems experiments are satisfactorily explained by the combined action of many bistable defects or traps. 
The distribution of the observed physical characteristics accounts for the rather large range of time constants generating $1/f$ noise.

Besides universality, other fundamental problems have been posed as the necessary existence of low-frequency cutoff to avoid unphysical divergencies and the related questions of stationarity. On the experimental side, 
it is often difficult to give an unambiguous interpretation of data as 
the ensemble-averaged power spectra provide 
incomplete information on individual fluctuators.

While for metals and spin glasses, some understanding of the mechanism of $1/f$ noise has been developed the problem is still unsolved for the majority of physical systems. Advances in the study of the microscopic nature are of practical importance since $1/f$ noise limits the performance of 
high-end electronic devices and sensors at low frequencies. Devices for metrological applications are also affected since their high operating frequencies show fluctuations with $1/f$ spectrum.

This chapter focuses on $1/f$ noise in \textit{quantum nanoscience}. Quantum nanoscience comprises of natural or engineered nanoscale systems whose functionality and structure can only be explained through quantum mechanisms such as discretisation, superposition and entanglement (definition from Ref. \citet{Milburn_2008}). These quantum properties facilitate, among other things, the enormous power of quantum information processing as well as novel quantum sensors. However, superposition states are fragile and quickly destroyed by $1/f$ noise, which needs to be considered for their functional applications. Our scope here is to discuss first the influence of $1/f$ noise on "traditionally-behaving" nanodevices and sensors, and then describe specific features encountered in the quantum coherent case.

New fundamental problems related to $1/f$ noise arise in quantum nanoscience and quantum information since the flicker noise may be the main source of decoherence present. For coherent quantum devices such as qubits, this has led to original ways for quantifying the impact of $1/f$ noise on the qubit, and to strategies for recovering the coherence using well-chosen pulse sequences. 

Progress in quantum nanoscience has also provided new tools for the characterization of $1/f$ noise sources. Nanoelectronic devices can be employed for example to detect charge noise at the micro electron charge level, thereby allowing to chart substrate characteristics in terms of trap states or fluctuators modelled by TLSs. Coherent TLS, qubits, on the other hand, can be employed as spectrometers detecting electromagnetic noise over a large span of frequencies. Qubits as spectrometers provide an excellent means to address the dynamics of impurity spins, which often form the culprit for coherence in quantum nanosystems.

\section{Theoretical overview}
\label{sec:overview}

For a stochastic process $x(t)$ gated in a time window $T$, $x(t) \neq 0$ for  $|t| < T/2$, the power spectrum $S_x(f)$, representing the spectral resolved ensemble-averaged power of the signal, is independent on $T$ if and only if the stochastic process is stationary.
For $1/f$-like noise the statistical properties change slightly over time
scales typically slow relative to the observation times. Thus it is considered a wide-sense (or weakly) stationary process, i.e. the correlation functions up to second order are (nearly) invariant for time translation. Then Wiener-Khintchine theorem applies, stating that $S_x(f)$ is given by the unilateral Fourier transform of the autocorrelation function
\begin{equation}
    \label{eq:autocorrelation}
\phi_x(\tau) := \langle \delta x(t+ \tau)\, \delta x(t) \rangle = \int_0^\infty \frac{d \omega }{ 2 \pi} \, S_x(\omega) \,\cos (\omega \tau ) \, .
\end{equation}
The integrated power in the spectrum between two frequencies $[f_1,f_2]$ is given by 
\begin{equation}
    \label{eq:inpowerspec}
    P_x(f_1,f_2) = \int_{f_1}^{f_2} \frac{d \omega }{ 2 \pi} \; S_x(\omega)
    = C_x \, \ln\Big(\frac{f_2}{f_1} \Big)
\end{equation}
implying that the integrated power per decade of frequencies is constant, a property known as scale invariance. The autocorrelation function for 
$\tau=0$ yields the variance $\langle (\delta x)^2 \rangle$ of the stochastic process which according to Eq.(\ref{eq:inpowerspec}) diverges logarithmically both at low and at high-frequencies. This divergence is known as a "paradox" of $1/f$ noise. Of course, the experimentally observed variance is finite since the frequency band for measurements is limited. Indeed it is bounded from below due to the finite duration of the realization of $x(t)$ being measured whose maximum value is set by the overall measurement time $t_m$. 
On the other hand, at large frequencies $1/f$ noise becomes much smaller than other contributions to fluctuations and it is not measurable, thus high frequencies are also cut off. This may happen at frequencies ranging from $\sim 10^2$ to $\sim 10^7$ Hz in different systems.

It is worth stressing that there is no experimental evidence for the existence of a low-frequency cutoff $f_c$,  therefore strictly speaking the stationarity of the process is still an open question. Difficulties are due to the fact that measuring the low-frequency part of the spectrum, say down to the sub-$\mu$Hz range would require a duration $t_m \sim 1/f_c$ of weeks. Therefore experimental statistical quantities depend on $t_m$. In particular this applies to the variance, $\langle (\delta x)^2 \rangle =   C_x \, \ln(f_2 t_m)$, which increases logarithmically with $t_m$. This is another peculiar property of $1/f$ noise of experimental relevance. In particular, it may have  a large impact on applications in coherent nanodevices where large fluctuations produce strong dephasing. From the mathematical point of view, this behavior of statistical quantities implies that $1/f$-like noise is not strictly stationary thus it is also non-ergodic. On the other hand, the logarithmic dependence on $t_m$ is weak enough to determine weak stationarity allowing the use of the Wiener-Khintchine theorem which relates the power spectra to correlation functions accessed experimentally. These latter show strongly nonexponential behavior decaying logarithmically with time in the important range $f_2^{-1} < t < t_m < f_c^{-1}$, another feature observed in several experiments. 
\begin{figure}
    \centering
    \resizebox{\columnwidth}{!}{\includegraphics{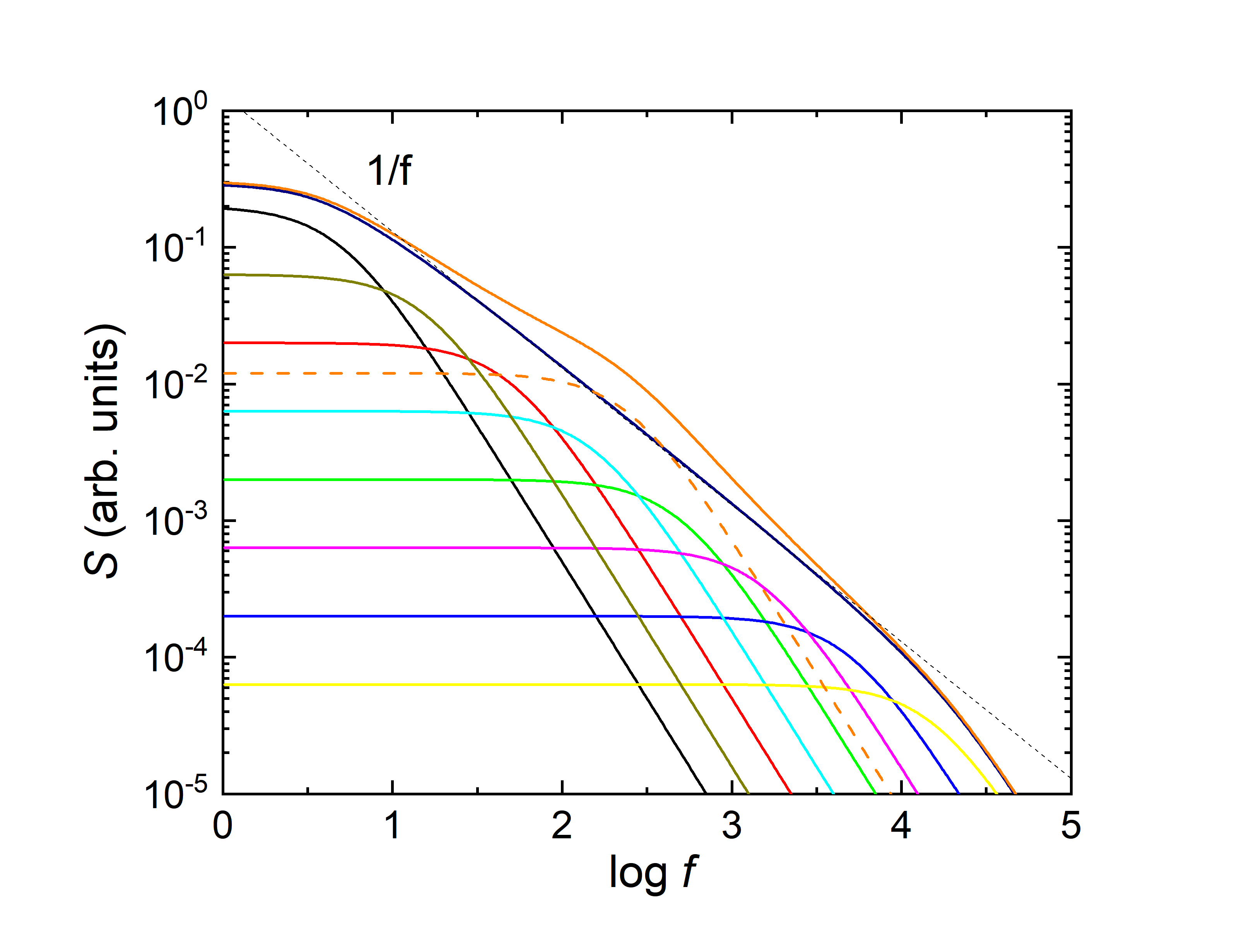}}
    \caption{
        Power spectral density of individual bistable fluctuators (coloured solid curves) whose combined effect (black solid curve) yields close to $1/f$ spectrum (indicated by dashed black line). If one of the fluctuators is more strongly coupled (dashed orange) a bump appears in the whole power spectrum (solid orange curve).}
    \label{fig:fluctuators}
\end{figure}

Understanding the microscopic nature of $1/f$ noise is a key problem both from the fundamental point of view and for applications aiming at fabricating improved electronic and coherent devices for modern nanoscience.  Typically, models for $1/f$ noise in the solid state are based on a collection of TLSs describing bistable defects or trap states with a wide distribution tunnelling rates~\citep{DuttaHorn_RMP,Kogan_1996,weissman1988RMP,Koelle_RMP_1999,Grasser2020}. A large collection of such states with broadly distributed parameters generates wide band $1/f$ noise and in what follows we will draw this  connection. 

Fluctuations of a bistable defect are described by a stochastic process $x_i(t) = \pm 1$ characterized by a single relaxation rate $\gamma_i$. Its kinetics leads to RTN where correlations decay exponentially, 
$\phi_i(\tau) = \exp (- \gamma_i \tau)$, yielding 
the Lorentzian spectral density 
$$
S_i(f) = 4 \int_0^\infty d \tau \,\phi_i(\tau)  \, \cos( 2 \pi f \tau) = 
\frac{4 \gamma_i}{(2 \pi f)^2 + \gamma_i^2}
$$
as given by the Wiener-Khintchine theorem. The overall noise is constructed as a linear superposition of these relaxation processes, $x(t)= \sum_i p_i x_i(t)$, and its power spectrum is given by
$$
S_x(f) = \sum_i p_i\, S_i(f) \,\to\, \int_{0}^{\infty} d \gamma \;
p(\gamma) \, \frac{4 \gamma}{(2 \pi f)^2  + \gamma^2}
$$
where we have allowed for a continuous distribution of relaxation rates. 
 Since $\langle (\delta x)^2 \rangle = \int_0^\infty df \, S_x(f) = \int_0^\infty d\gamma \, p(\gamma)$,  the quantity $p(\gamma)\,d \gamma$ 
 is the contribution to the variance of processes with rates in the interval $[\gamma,\gamma + d \gamma]$. For a weighting function $p(\gamma) = C_x/\gamma$ in the interval $[\gamma_1,\gamma_2]$ and vanishing 
 elsewhere we obtain 
 
\begin{equation}
 \label{McW}
 S_x(f) =  \int\limits_{\gamma_1}^{\gamma_2} \frac{ 4 C_x \,d \gamma}{( 2 \pi f)^2 + \gamma^2} = \left\{
 \begin{array}{ll}
\mbox{const} &\; f \ll \frac{\gamma_1 }{ 2 \pi}
\\[2pt]
{C_x/ f} &\; \frac{\gamma_1 }{ 2 \pi} \ll  f \ll \frac{\gamma_2 }{ 2 \pi}
\\[2pt]
\frac{C_x (\gamma_2 -\gamma_1) }{ (\pi f)^2} 
&\; \frac{\gamma_2 }{ 2 \pi} \ll f
 \end{array}
 \right.
\end{equation} 
reproducing the $1/f$ behavior for intermediate frequencies,  
 $\gamma_1 \ll 2 \pi f \ll \gamma_2$ (see Fig.~\ref{fig:fluctuators}). 
 Notice that the extremal rates $\gamma_1$ and $\gamma_2$ provide soft frequency cutoffs to $S_x(f)$, avoiding the divergence of $\langle (\delta x)^2\rangle$. 
 They are easily identified for a finite system of discrete fluctuators. We have still to justify the form of $p(\gamma)$ which is the other key assumption determining the $1/f$ behavior of the power spectrum. In the Mc Worther model~\cite{McWorther1957} for semiconductors, the kinetics is modelled as tunneling processes. The rates have the form $\gamma(z) = \gamma_0 \,\exp(- z/C)$ depending on the parameter $z$ summarizing the exponential dependence on the width and on the height of the tunnel barriers. Now, if $z$ is distributed uniformly in the interval $[z_2,z_1]$ then  $p(\gamma)=C/\gamma$ in $[\gamma(z_1),\gamma(z_2)]$. The same result is obtained if the kinetics stems from thermal activation, $\gamma(E) = \gamma_0 \,\exp(- E/k_B T)$, assuming 
 that the activation energies $E$ are uniformly distributed as originally proposed by Du Pré and by Van der Ziel~\cite{DuPre1950,vanDerZiel1950}.  Notice that this picture naturally accounts for the weak stationarity (implying non-ergodicity) of $1/f$ noise since during the measurement, the TLSs with relaxation rates $\gamma < 1/t_m$ are frozen in some non-equilibrium configuration.

 Specific models for $1/f$ noise based on quantum TLSs interacting with phonons (in insulating solids) or electrons (in metals) yielding physically motivated distributions of parameters have been used to explain observed properties of $1/f$ noise. These models apply in rater simple situations of non-interacting impurities while in general, the physical properties of disordered systems depend on their many-body nature. Still, the $1/f$ behavior in the solid-state appears to be a consequence of the universal properties of the kinetics rather than the universality of low-frequency fluctuations~\citep{Kogan_1996}. 

Finally, we remark that noise from a collection of TLSs is not Gaussian. The Gaussian limit is approached  
when the noise is generated by a sufficiently large number of independent fluctuators each of which contributes negligibly to the variance.  Non-Gaussianity in $1/f$ noise has been observed
in different nanosystems, although
it poses experimental challenges since it requires finite-bandwidth measurement with large observation times $t_m$~\citep{weissman1988RMP}. On the other hand, resolving the contribution of individual fluctuators 
is needed in view of miniaturization and of the study of the impact of $1/f$ fluctuators on the reliability of quantum architectures. Non-Gaussian features of $1/f$ noise and their impact
on phase-coherent quantum systems are discussed in depth in Ref. \citet{Paladino2014}.

\section{1/f noise in nanoscience}
The relative $1/f$ resistance fluctuations $\delta R /R$ have been found to decrease as $1/\mathrm{Volume}$ in bulk materials. Similarly, the relative $1/f$ current noise in FETs has been observed to scale as $\propto 1/\mathrm{Area}$ of the gate electrode. So inevitably, the role of $1/f$ noise will grow when going to nanoscale quantum devices. Lately, nano-scale FETs have been fabricated using heterostructures of atomically thin 2D materials. In such components, the surface-to-volume ratio becomes the largest, which may lead to major issues with the quality of surfaces or interfaces in such nanodevices. As Wolfgang Pauli said "God made the bulk; surfaces were invented by the devil", this statement becomes more and more relevant when entering the nanoscale regime.

Quantized conductance in nanoscience (mostly 1D or 2D transport) is typically described using the Landauer-B\"uttiker formalism \citep{Datta1997}, in which current is given as a sum over transmission channels with transmission $\tau_{ik}$: $I=\Sigma \tau_{ik} G_0$ where $G_0=2e^2/h$ and the indices $k$ and $i$ refer to subbands and quantized states within them, respectively. If we take only one channel with transmission $\tau$, the conductance can be written as $G^{-1}= (1-\tau)/\tau + G_0^{-1}$ where the first term is the four-terminal resistance and the second term describes the contact resistance for perfect contacts. Since contact resistance is constant, fluctuations in $\tau$ yield $\delta R/R=- \delta \tau/\tau$ both for the two- and four-lead resistance $R$. If the fluctuations in transmission channels are independent, we may add fluctuations of the channels incoherently, which yields $\delta G^2=\Sigma  \delta \tau_{ik}^2 G_0^2$. Each fluctuating channel can have its own frequency dependence, which in the simplest case is Lorentzian with a cut-off given by a characteristic two-level fluctuator frequency $f_{ik}^c$ (see below). If the cut-off frequencies $f_{ik}^c$ are distributed uniformly on log-scale, summation over all $\delta \tau_{ik}(f)$ yields a $1/f$ noise spectrum in transmission. In practice, contacts are not ideal and may also have fluctuations. In the nearly ballistic limit, the nonidealities of contacts can be integrated into the transmission values $\tau_{ik}$, and no distinction between scattering in the contact vs. the channel needs to be made. 

The conductance in diffusive electrical transport can be written as $G=eN\mu$ where $N$ is the total number of charge carriers and $\mu $ is their mobility; $e$ is the electron charge. Therefore, the conductance variation is either due to mobility or carrier number fluctuations:
\begin{equation}
\label{deltaG}
\delta G = eN\delta \mu  + e\mu \delta N
\end{equation}

 In bulk metallic conductors, $N$ can be made very large, and the relative fluctuations coming from carrier number can be made to vanish as $1/\sqrt{N}$. However, when the sample size is brought towards the nano regime, the dominance of mobility fluctuations is suppressed and both contributions have to be taken into account. Carrier number is varied by tunneling into trap states that reside in the nearby dielectric, for example in the gate dielectric of a MOSFET transistor. In a small quantum device, such as a quantum point contact (QPC), the trap state may result in distinct RTN signature in the resistance (see Sec. \ref{sec:overview}).
 The RTN noise in a QPC can be described using Machlup's formula for a single two-level fluctuator \citep{Machlup1954}:
\begin{equation} 
\label{Machlup}
{S_G}(f) = {\left[ {G_{\tau_1} - G_{\tau_2}} \right]^2}\frac{{{w_1}{w_2}\tau }}{{1 + {{(2\pi f\tau )}^2}}} 
\end{equation} 
where $w_1$ and $w_2=1-w_1$ denote the probabilities for finding the QPC in conductance states $G_{\tau_1}$ and $G_{\tau_2}$, respectively, while $1/\tau$ denotes the total transition rate between the conductance states (sum of back and forth rates).

Even though nanoscience devices are small, more than one fluctuator is typically acting on them. The McWhorter model is obtained by considering a collection of single fluctuators which have an equal strength and a uniform distribution of corner frequencies $f_c=1/(2\pi \tau)$ on logarithmic scale. The resulting $1/f$ spectrum is given in Eq. (\ref{McW}) where $1/\tau$ corresponds to $\gamma$.

This $1/f$ spectrum will cover a large frequency span provided that the trap lifetimes $[\tau_1 \dots \tau_2]$ are distributed over a wide range. In the standard McWhorter model,  the trapped charge is set to one electron charge $e$, but with separate TLSs, the charge change in the device induced by the fluctuators may be a fraction of $e$, which reduces the prefactor $C_x$ in Eq. (\ref{McW}).  The McWhorter model has been  successfully applied to semiconductor devices and, indeed, good agreement is obtained between experimental results in MOSFETs  and the model \citep{Burstein1957,Grasser2020}.

Typically, the noise is given as relative fluctuations 
${\left( {\frac{{\delta G}}{G}} \right)^2} = {\left( {\frac{{\delta R}}{R}} \right)^2} = {\left( {\frac{{\delta N}}{N}} \right)^2} + {\left( {\frac{{\delta \mu }}{\mu }} \right)^2}$, i.e. scaled by the total conductance of the system.
Ideally, the scaling conductance should be governed by those processes which are governing the fluctuations. Unfortunately, this is not always the case and then the relevant scattering processes should be separated from the data before scaling.  Theories can also incorporate correlations between carrier number and mobility fluctuations. The fluctuations are correlated because an electron in the trap state can also modify the scattering governing the mobility of the sample. This is common in 2D material FETs, for example, fabricated using graphene \citep{Balandin2013}.

Good electrical contacting is difficult to achieve in nanosamples. Consequently, the low-frequency noise formula in Eq. (\ref{deltaG}) has to be supplemented by the contact noise term, which adds incoherently with the noise arising from the nanodevice. For the relative fluctuations in the total conductance $G_T$ one obtains then
$
{\left( {\frac{{\delta G_T}}{G_T}} \right)^2} = \frac{{{R_c^2}}}{{{{(R + {R_c})}^2}}}{\left( {\frac{{\delta R_c}}{R_c}} \right)^2} + \frac{{R^2}}{{{{(R + {R_c})}^2}}}{\left( {\frac{{\delta R}}{R}} \right)^2}
$
in which $R=1/G$ and $R_c$ denote the sample and contact resistance, respectively, and the weighting factor of the relative contact noise ${\frac{{\delta R_c}}{R_c}}$ grows with the ratio $\frac{{{R_c}}}{{(R + {R_c})}}$.

Both $\delta \mu$ and $\delta N$ in Eq. (\ref{deltaG}) may depend on temperature, which can lead to low-frequency conductance fluctuations owing to temperature fluctuations $\delta T$ \citep{Kogan_1996}.
The influence of $\delta T$ fluctuations is expected to increase in small nanodevices, because $\left< \delta T^2 \right> \propto k_B T^2/C$ where $C$ is the heat capacity of the system. These temperature fluctuations can, for example, limit the ultimate energy resolution in 
 nanosized calorimeters, as discussed in Ref. \citet{Karimi2020}.

TLSs discussed in Sec. \ref{sec:overview} lead to noise and dissipation both in acoustic and electric systems. The acoustic coupling is due to strain fields (elastic dipole moment) that every TLS possesses when embedded into elastic material \citep{Lisenfeld2019}.  Some of the TLSs possess also an electric dipole moment, which makes them a source of $1/f$ noise in electrical circuits, both for the conductance noise as well as for fluctuations in dielectric properties of materials \citep{Zmuidzinas2012,Lisenfeld2019}. Therefore, TLSs make a contribution to the real and imaginary parts of the dielectric constant ${\varepsilon_{TLS}(f,T)}$. In weak  electric fields, the loss tangent $\delta_{TLS}(f,T)$ of the dielectric constant is dominated by TLS energies close to $hf$:
\begin{equation} \label{loss}
{\delta _{TLS}}(f,T) = \frac{{\Im {\varepsilon_{TLS}(f,T)}}}{{\Re{ \varepsilon}}} = {\delta_0}\tanh \left( {\frac{{hf}}{{2{k_B}T}}} \right) 
\end{equation}
 where the hyperbolic tangent describes the equilibrium population difference of the two levels and $\delta_0$ is the intrinsic loss parameter proportional to the density of TLS per unit volume and energy. If the asymmetry and tunneling parameters of two-level states are uniformly distributed in energy as assumed in the standard tunneling model \citep{Lisenfeld2019}, the low-temperature loss tangent becomes independent of frequency. 
In the limit $k_B T \gg hf$, the upper state will become populated, leading to $\delta_{TLS}(f,T) \propto hf/k_B T$. Clearly, the role of TLSs in dissipation vanishes when $f \rightarrow 0$. According to Kramers-Kronig relations, Eq. (\ref{loss}) can be utilized to obtain the reactive part of $\varepsilon_{TLS}(f,T)$, and the equations for fluctuations of reactance (capacitance) can be derived \citep{Zmuidzinas2012}. 

The capacitance modulation due to TLSs provides an efficient means to investigate two-level fluctuators and the ensuing $1/f$ noise using superconducting microwave cavities \citep{Zmuidzinas2012,Lisenfeld2019,Pappas2020}. Furthermore, understanding the performance limits of such cavities is vital, because high-quality superconducting microwave cavities are critical elements for efficient qubit readout \citep{Blais2021} and kinetic inductance detectors for sensitive bolometry or single photon detection \citep{Ulbricht2021}. 
A summary of recent results obtained for the loss tangent are given in Ref. \citep{Pappas2020} for several dielectric and superconducting materials which are relevant  for coherent quantum nanoscience. Most recently, promising results have been obtained for superconducting tantalum \citep{Place2021}; in comparison to Nb, this is assigned to more kinetically limited and chemically robust oxides in Ta \citep{deLeon2021}.

Measurement power is known to modify the influence of TLSs \citep{Pappas2020}. Dissipation in a superconducting material is first governed by TLSs, but increasing power leads to the creation of quasiparticles that dominate losses at high rf power. On the other hand, when there are spatial variations in the superconducting order parameter, i.e. the superconducting gap, it is possible to have quasiparticles trapped into local minima of this gap variation. In fact, TLSs  due to trapped quasiparticles on the surface of Al superconductor have been observed to produce low energy TLSs up to energies of $\mathrm{h} \times 20$ GHz \citep{deGraaf2020}.  The observed level separation agrees with theoretical values obtained  using estimates of the second derivative of the spatial variation of the gap \citep{deGraaf2020}.

In standard TLS models, the number of two-level states $N$ is evenly distributed in energy, i.e. the density of states $\mathfrak{N}(E)=dN/dE = \mathrm{const.}$ Plenty of experiments have been performed to test this hypothesis \citep{DuttaHorn_RMP}. In quantum nanodevices, there are recent experiments that indicate an energy-dependent density of states $\mathfrak{N}(E)$ \citep{Lisenfeld2019}. Energy dependence can arise for example from interactions between TLSs, which yields an Efros-Shklovskii type of pseudogap at small energies. There are indications that $\mathfrak{N}(E) \propto E^{\alpha/2}$ with $\alpha \sim 0.2-0.7$; the same exponent appears also  in the temperature dependence of $1/f$ noise $S_V \propto 1/T^{\alpha}$ \citep{Burnett2014,Lisenfeld2019}, which leads to increase of $1/f$ noise with decreasing $T$.

Surface spins, such as physisorbed atomic hydrogen, have been found to behave as TLSs that cause significant dielectric and flux noise. Removal of surface adsorbants by suitable physical or chemical treatment has led to a factor of 10 reduction in frequency fluctuations of a microwave cavity at $\sim 5$ GHz frequency \citep{deGraaf2018}. 
The scaling of $1/f$ as $\propto \mathrm{Area}$ is quite conspicuous in superconducting quantum nanoscience, for example in the superconducting loops SQUIDs (Superconducting Quantum Interference Devices, see Sec. \ref{noise_QC}). It is also becoming apparent that surfaces and interfaces impose large noise contributions in superconducting microwave circuits, which may even dominate the defect/impurity contributions scaling with volume. 

Besides TLSs, non-equilibrium quasiparticles may cause excess noise and losses in superconductors. In general, the number of quasiparticles observed in superconducting films at $T \lesssim 150$ mK amounts to $10^{-8} \dots 10^{-6}$ per Cooper pair, which is clearly larger than predicted from the BCS theory for a superconductor in equilibrium with the environment at $T=10$ mK. The reason for excess quasiparticles is not fully resolved yet, but recent experiments have indicated that non-equilibrium phonons, created by e.g. cosmic rays, do play an essential role in the process \citep{McEwen2022}.

\subsection{Noise in quantum circuits} \label{noise_QC}
Quantum circuits are based on two canonically conjugate variables, charge $Q$ and flux $\Phi$ (see Sec. \ref{AA}).
The strength of quantum fluctuations in these variables depends on the impedance of the quantum circuit $Z_0$. If $Z_0$ is large (fluctuations for the impedance $Z_0=\sqrt{L/C}$ are derived from corresponding LC oscillator \citep{Devoret1997}), the fluctuations of charge are given at the quantum limit by $\left< \delta Q^2 \right> = \frac{\hbar}{2Z_0}$, while the fluctuations for flux amount to $\left< \delta \Phi^2 \right> = \frac{\hbar Z_0}{2}$. Hence, quantum circuits at large $Z_0$ are susceptible to external charge noise while external flux noise is detrimental in the case of small $Z_0$.

Flux noise is produced by resistance fluctuations when current is passed through a normal conductor. In metallic conductors, the commonly accepted view is that $1/f$ noise is determined entirely by fluctuations in charge carrier mobility \citep{Hooge1981a}. In addition to models based on localized states and fluctuating scattering cross sections \citep{Hooge1994}, there are more refined models based on a modification of electronic wave interference induced by a defect displacement by a hopping event. These models, the universal conductance fluctuation noise and local interference noise model, have been quite successful in accounting for many experimental observations in metals with defects as summarized in Ref. \citet{Giordano1989}.

\vspace{6pt}
\begin{center}
\emph{\small{Superconducting Quantum Interference Devices}}
\end{center}
\vspace{2pt}
Classical SQUIDs are inherently low impedance devices, which is highlighted by the small shunt resistance inserted in them for stabilization of their phase dynamics. Hence, their ultimate sensitivity is limited by the low-frequency flux noise, or alternatively by internal fluctuations of the Josephson energy of the tunnel barriers.
Low-frequency noise in SQUIDs displays distinct $1/f^{\alpha}$ behavior \citep{Wellstood1987,squid2004}.  In spite of 50 years of investigations, the origin of flux noise in SQUIDs is still poorly understood. It has been found that its power law exponent 
$\alpha$ varies substantially around $\alpha = 1$, and the exponent displays strong temperature dependence \citep{Anton2013}.
An increase of noise linear with the area of the superconductor in the SQUID loop has been found \citep{Anton2013}, but no clear correlations have been observed neither on the selection of materials nor on the details of the device geometry. Even though the fundamental mechanism for flux noise in SQUIDs remains unclear, there are plenty of experiments indicating that impurity atoms with a spin magnetic moment are involved in the noise phenomena.

Flux noise in a SQUID may originate from paramagnetic impurities residing on the superconducting material of the SQUID loop.  Paramagnetic impurities produce flux according to paramagnetic susceptibility $\chi(T)  \propto 1/T$, the fluctuation of which produces flux noise. Such fluctuating paramagnetic impurities have been observed in Al, Nb, Au, Re, and Ag among others
\citep{Sendelbach2008,Bluhm2009}. The fluctuating spins are believed to reside on the metal surface as the flux noise scales with the surface area of the SQUID loop \citep{Anton2013}. In order to obtain a correct order of magnitude of flux noise, one has to assume a magnetic impurity concentration of $\sim 1\times10^{18}$/m$^2$  with an impurity moment of one Bohr magneton, given by electron rest mass $m_e$ and electron charge $e$ as $\mu_B=\frac{e \hbar}{2m_e}= 9.274 \times 10^{-24}$ J/T, i.e. a distance of 1 nm between the impurities \citep{Sendelbach2008}. There is also experimental evidence for adsorbed oxygen molecules producing a magnetic signature and flux noise \citep{Kumar2016}. With its paramagnetism, oxygen can meet the above-stated requirement for strength and density of impurity moments.

Hydrogen atoms provide an alternate source for flux noise \citep{deGraaf2017,Quintana2017}. Electron spin resonance measurements on sapphire have revealed a resonance with a 1.42 GHz line splitting which matches free H atoms. In addition, flux noise measurements using the qubit relaxation rate indicate a peak at 1.4 GHz. Further support for the presence of H is provided by DFT calculations of interstitial hydrogen atoms embedded in bulk sapphire, which yield an ESR splitting of $1.28 - 1.36$ GHz \citep{Wang2018a}. Recently, complex radicals have also been found to contribute to the noise and decoherence phenomena \citep{Un2022}.

\vspace{6pt}
\begin{center}
\emph{\small{Single electron tunneling devices}}
\end{center}
\vspace{2pt}
The most sensitive charge detectors in the solid state are single electron transistors (SET) and charge qubits. Both devices suffer strongly from background charge fluctuations which often have $1/f$ character with some overlying Lorentzian spectra due to strongly coupled TLSs \citep{Paladino2014}. Using Al/AlO$_x$/Al tunnel barriers on top of common dielectric such as Si, SiO$_2$, and Si$_3$N$_4$, the charge noise amounts to $10^{-4}-10^{-3} \textrm{e}/\sqrt{\textrm{Hz}}$ at 10 Hz, both in superconducting and normal metal regimes. The best Al/AlO$_x$/Al devices using stacked design have reached $8 \times 10^{-6} \textrm{e}/\sqrt{\textrm{Hz}}$ at 10 Hz \citep{Krupenin2000}. Charge noise in SETs arises typically from charged two levels states capacitively coupled to the island \citep{Paladino2014}. Such TLSs are present in all back-gated devices at the interface of the gate dielectric and the conducting materials. The best charge sensitivities in Al-based devices have been obtained in structures where the contact area to dielectric substrate material has been eliminated with a stacked design \citep{Krupenin2000}. The achieved levels of charge noise have not been adequate for fabricating well-functioning charge qubits, i.e. Cooper pair boxes, using Al-technology \citep{Astafiev2006,Paladino2014}.

$1/f$ charge noise is a very important characteristic for industrially fabricated silicon quantum dot MOSFET devices which are being developed for spin qubits \citep{2022Harvey}. Typical noise values in multielectron quantum dots at 1 Hz are in the range $1-10$ $\mu$V referred to the gate electrode \citep{Petit2018,Zwerver2022} (corresponding to $80-800 $ $\mu \mathrm{e}/\sqrt{\mathrm{Hz}}$ using a gate capacitance $10^{-17}$ F). Small charge noise in these quantum dots is important both for the coherence and read-out of spin qubits \citep{Stano2022}. 

In 1D and 2D nanodevices, the density of states can be strongly energy dependent. This causes fluctuations in the chemical potential $\mu_F$ to generate carrier number fluctuations, which leads to a one-to-one correspondence between the conductance noise spectrum and the frequency spectrum of $\mu_F$ fluctuations. The fluctuations in $\mu_F$ may be induced, for example, by chemical doping or background charge fluctuations \cite{Pellegrino2019}. Such fluctuations in 1D and 2D conductors are similar to Coulomb effects, where the shift of chemical potential is given in terms of variation in the island charge  $Q=Q_{\mathrm{i}}$. The influence of $Q_{\mathrm{i}}$ on the current $I$ can be described using transconductance $g_m=\partial I/\partial V_g=C_{eff}\partial I/\partial Q_{\mathrm{i}}$ where $C_{eff}$ specifies the connection of the island charge to the shift of electrochemical potential. Typical charge noise values for carbon nanotube \citep{Roschier2001,Vitusevich2011} and 2D material devices \citep{Karnatak2017} with good transconductance are in the range $4 \times 10^{-5} \dots 10^{-3} \textrm{e}/\sqrt{\textrm{Hz}}$ at 10 Hz. 

Besides charged TLSs, background fluctuations can arise from charge localized at interfaces where it can be transported by impurities/defects easier than within the bulk. This calls for extreme care when stacking 2D heterostructures in order to keep their $1/f$ noise as small \citep{Balandin2013,Karnatak2017}. In addition, there has been a substantial effort in studies on suspended 2D materials, in particular, graphene, which has been employed as a low-noise platform for investigations of noise induced by diffusion of mobile scattering centers on the membrane \citep{Kamada2021}. As an interpretation of the observed noise results, intrinsic long-term correlations due to clustering and declustering of adsorbed atoms has been proposed \citep{Kamada2021}. Interactions between TLSs and non-Gaussian fluctuations can also be investigated by the second spectrum \cite{Pellegrino2023,Zeng2023}.

\section{1/f noise in quantum-coherent nanodevices}
\label{sec:oneoverf-coherent}

Nanoscale-based quantum technologies build on harnessing fundamental properties of quantum states, such as superposition, quantum coherence and entanglement
to enable novel functionalities \citep{Heinrich2021}. The extraordinary impetus in quantum-technology research, from academia to spin-offs and industries, since the last few years of the 20$^{\rm th}$ century,    
has led to a number of breakthroughs. At the same time, it has pointed out that  fluctuations characterized by power spectral density  $1/f^\alpha$ with $\alpha \approx 1$ in a setup-dependent low-frequency range,
are paramount in most of the coherent nanodevices. On one side, they limit the  ability to control the quantum-coherent behavior of 
engineered nanostructures, leading to a  gradual loss of phase coherent evolution or  \textit{decoherence}. 
On the other side, the systematic analysis of decoherence under different control sequences allows for reconstructing the frequency spectrum of stochastic signals, a task commonly referred to as \textit{noise spectroscopy}.
Here, we present an overview on how fluctuations with $1/f$ spectrum influence the phase coherent evolution of artificial atoms playing the twofold roles of elementary quantum bits and of quantum sensors, in particular of noise spectrometers. We will refer to selected paradigmatic examples of coherent nanodevices.

\subsection{Artificial atoms} \label{AA}

Nowadays micro and nano solid-state devices can be fabricated. Tehy are called {\em artificial atoms} (AA)~\citep{Buluta2011}  since they show functional analogies with natural atomic or molecular systems. 

Electrons in semiconductors can be confined in a  "box" called {\em quantum dot}, small enough that they do not behave as in bulk solids but rather as electrons in individual atoms. Indeed they have a discrete spectrum of energy and at low temperatures, electrons can occupy only the lowest-energy levels. A qubit can be encoded in the charge or the spin of electronic configurations. Spin qubits couple weakly to their environment and leveraging on the available fabrication technologies they are interesting candidates for building large-scale quantum computers. 

\begin{figure*}[t!]
\resizebox{0.95\textwidth}{!}{\includegraphics{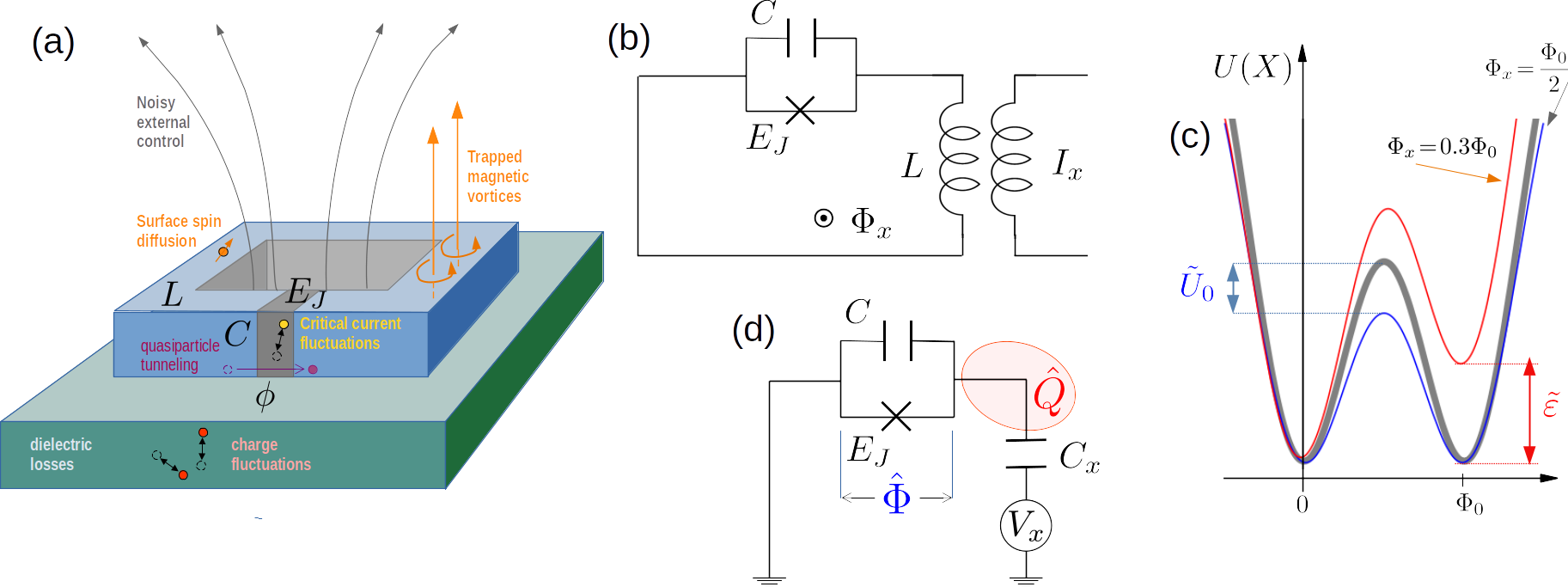}}
\caption{(a) The rf-SQUID is a superconducting ring with inductance 
$L$, and associated energy $E_L = \Phi_0^2/L$, interrupted by a Josephson junction with critical current $I_c$, Josephson energy $E_J = 2 e \, I_c/\hbar$ and capacitance $C$; several noise sources may affect its dynamics, many of them displaying a $1/f$ low-frequency behavior. (b) The equivalent circuit of the rf-SQUID, the flux $\Phi_x$ being induced by a current $I_x$ flowing in a nearby circuit. (c) 
The rf-SQUID potential $U(X)$ Eq.(\ref{eq:rfSQUID-H}): the device 
is analogous to a particle of "mass" $C$ moving in this potential, modulable by varying $\Phi_x$. (d) Equivalent circuit of the Cooper-pair box, where a discrete-valued charge $\hat Q$ is stored in a superconducting island (red) connected to the circuit by a Josephson junction and modulated by the external voltage $V_x$ via a gate capacitance $C_x$.
\label{fig:artificialatoms}
}
\end{figure*}

Superconducting artificial atoms are devices made of tunnel junctions displaying the Josephson effect. In a superconductor, the many-body condensate is described by a single-particle wavefunction, and in coherent devices, its phase behaves as a quantum degree of freedom analogous to the coordinate of a fictitious particle in a confining potential. Superconducting AAs may encode a qubit in the eigenstates of the potential (phase qubits), in the flux enclosed in a superconducting quantum interference device (SQUID) ring (flux qubit), in the extra charge stored in a superconducting "box" (charge qubits) to mention the simplest instances (see Fig.~\ref{fig:artificialatoms}).

As an example, the rf-SQUID illustrated in Fig.\ref{fig:artificialatoms}a is modelled by an equivalent circuit with concentrated elements (Fig.~\ref{fig:artificialatoms}b). The circuit's dynamics emulates a particle moving in the potential
\begin{equation}
\label{eq:rfSQUID-H}
U(X)= - E_J \, \cos \frac{2 \pi X }{ \Phi_0} + \frac{E_L }{ 2} \,\bigg(\frac{X - \Phi_x }{ \Phi_0}\bigg)^2 
\end{equation}
where the "coordinate" $X$ is the total magnetic flux concatenated with the ring $\Phi_T$ and the "momentum" $P$ is the charge $Q$ accumulated at the surface of the junction's electrodes. The Hamiltonian describing the quantum-coherent regime is obtained by quantizing the canonical coordinates which become non-commuting operators, $[\hat Q, \hat \Phi_T]=i \hbar$. Here
$\Phi_x$ is the flux of an external bias field concatenated to the ring, 
$\Phi_0= h/(2e) \approx 2.068 833 \times 10^{-15} \mathrm{Wb}$ is a universal constant named \emph{flux quantum}  and other circuit parameters are defined in 
Fig.~\ref{fig:artificialatoms}).

In principle, an rf-SQUID biased by an external flux $\Phi_x \approx  \Phi_0/2$ may function as a qubit. In this case $U(X)$ is a nearly symmetric double-well potential (see Fig.~\ref{fig:artificialatoms}c) with a low-energy spectrum consisting of a doublet. The qubit Hamiltonian is obtained by truncating the system to these two levels. In the spin language, it is written as
\begin{equation}
    \label{eq:qubit}
    \hat H = - \frac{\varepsilon(\Phi_x) }{ 2 }\, \sigma_z - \frac{\Delta }{ 2 }\, \sigma_x 
\end{equation}
where $\varepsilon$ is the asymmetry of the potential, modulable by $\Phi_x$
and $\Delta$ is a tunneling matrix element between its minima. The two eigenvalues of $\sigma_z$ correspond to states localized in one of the minima of the potential, the total concatenated flux being $\Phi_T =0,\Phi_0$. These are the basic principles of the so-called flux-qubit whose actual design is a multijunction version of the rf-SQUID \citep{Krantz_2019}. 

Classical noise induces fluctuation of the parameters. For instance, flux noise is described by adding a stochastic external flux, $\Phi_x \to \Phi_x + \tilde \Phi_x(t)$, which induces fluctuations $\tilde \varepsilon(t)$ of the potential asymmetry, as illustrated in Fig.~\ref{fig:artificialatoms}c. Instead, fluctuations of $E_J$ due for instance to the impurities in the Josephson junction induce fluctuations of the barrier height $\tilde U_0$ of the potential and in turn fluctuations $\tilde \Delta(t)$ of the tunnel splitting. In general, the Hamiltonian acquires a fluctuating term $\delta \hat H(t) = - \frac{1 }{ 2} \big[ \tilde \epsilon(t) \,\sigma_z + \tilde \Delta(t)\,\sigma_x \big]$ where $\tilde \epsilon(t)$ and $\tilde \Delta(t)$ are classical stochastic processes. Experiments show that all of them exhibit $\sim 1/f$ behavior at low frequencies~\citep{Paladino2014,bylander2011NatPhys}. The approach outlined for classical noise may be extended to describe phenomenologically quantum noise. In this case, the stochastic processes are substituted by operators describing physical fields exerted by the microscopic quantum degrees of freedom of one or more physical environments~\citep{Paladino2014}.

The other prominent family of superconducting qubits, including the transmon~\citep{Koch2007} which provides the forefront device for present superconducting quantum processors~\citep{Arute_2019,ibm2021,china2021}, 
may be introduced following the same roadmap. The simplest instance 
is the so-called charge qubit, Fig. 3(d) ~\citep{Nakamura_1999} where the conjugate variables are the discrete charge $\hat Q$ into the superconducting island and the cyclic coordinate $\hat \Phi= \hbar \hat \phi/(2 e)$ related to the 
 quantized phase difference $\phi$ between the macroscopic wave functions of the electrodes of the Josephson junction.  
Again the Hamiltonian can be written in the form Eq.(\ref{eq:qubit}) where now fluctuations $\tilde \varepsilon$ are related to charge noise 
whereas $\tilde \Delta(t)$ originates from fluctuations of $E_J$ or from flux noise.

 The main effect of low-frequency noise on the dynamics of the qubit is dephasing. For instance, the populations of the eigenstates of an  isolated qubit with Hamiltonian Eq.(\ref{eq:qubit}) undergo coherent oscillations with the angular frequency given by the qubit energy splitting (we take $\hbar=1$) $\Omega=\sqrt{\epsilon^2 + \Delta^2}$. 
Due to the coupling to the environment, which induces fluctuations of $\Omega$,
the amplitude of coherent oscillations decays as $A(t)= \mathrm{e}^{- \Gamma(t)}$. 
Dephasing strongly depends on how the qubit is biased or manipulated. This extreme sensitivity is exploited to mitigate dephasing or, on the contrary, for quantum sensing of environmental noise. For instance, let's set the flux bias such that $\varepsilon(\Phi_x)=0$ and consider the effect of fluctuations $\tilde \Delta$ (called "longitudinal"). Then  the decay of oscillations is given by 
\begin{equation}
\label{eq:gamma-long}
\Gamma(t) =\frac{t^2 }{ 2} \int_{t_m^{-1}}^\infty d f \; {\cal F}_0(ft) \, S_\Delta(f)
\end{equation}
being determined by the power spectrum of the fluctuations of $\tilde \Delta(t)$ "filtered" by the function ${\cal F}_0(x) = \sin^2(\pi x)/(\pi x)^2$. Longitudinal $1/f$ noise may determine a very strong decay $\Gamma(t) \approx \frac{1 }{ 2} \langle (\delta \tilde \Delta)^2\rangle t^2$ where $ \langle (\delta \tilde\Delta)^2\rangle \sim 
\int_{1/t_m}^{f_2} d f \, C_\Delta/f \approx  C_\Delta \,\log (f_2 t_m)$. This strong decay of coherences determines the complete phase randomization in a generic solid-state system at equilibrium. Instead, $\tilde \varepsilon$ noise (called "transverse") contributes at second order on the fluctuations of the qubit splitting $\Delta$ thus 
their impact is more limited resulting in a power-law decay
\begin{equation}
 \label{eq:gamma-transv}
 \Gamma(t) = \frac{1 }{ 4} \, \ln\Big[ 1 + \Big(\frac{\langle (\delta \tilde \varepsilon)^2\rangle \, t }{ \Delta}\Big)^2 \Big]
\end{equation}
Therefore, to reduce the impact of low-frequency noise, an optimal bias strategy has been used, where the qubit is biased in such a way that the main source of noise is transverse. This way, the impact of flux (charge) noise in flux (Cooper-pair box-based) qubits is minimized. 

In the last generation of qubits, significant coherence times are achieved by designing the device in a way that protection from noise is stronger for a larger range of parameters. 
For instance, dynamical sweet spots based on the application of periodic drives have been identified \citep{Huang2021}, demonstrating reduced susceptibility to low-frequency noise 
and predicting an enhancement of pure dephasing by 3 orders of magnitude
in the fluxonium \citep{Nguyen2019}.
For a recent review on noise-protected superconducting quantum circuits, we refer the reader to Ref. \citet{GyenisPRXQuantum2021}. Here we mention that, in addition to design strategies, considerable advances in materials science and engineering materials provided significant noise protection, as discussed in Ref. \citep{deLeon2021}. 

In general,  amorphous or glassy materials with plenty of TLSs may be disadvantageous in coherent quantum devices. Besides residing in the tunnel barrier material, TLSs may influence the dielectric constant of substrates as well as insulating coatings between wiring layers. The inverse time constants of the two-level states have been found to range from mHz to GHz frequencies. Hence,  in addition to limiting the quality factors of qubits, TLSs can influence readout resonators for qubits at
GHz frequencies as well as low-frequency SQUID magnetometers, as discussed in the next paragraph. 

Defects in oxides and interfaces also represent the main limitation of gate fidelities in semiconductor-based quantum 
dot charge qubits, where they cause voltage fluctuations on the control electrodes. Analogously to superconducting qubits,  dephasing due to $1/f$-charge noise is minimized by working at sweet spots \citep{Petersson2010,Dial2013,Thorgrimsson2017,Scarlino2019,Yang_2019,Kratochwil2021}.
Spin-based quantum dot qubits \citep{1998Loss} in semiconductors, 
are sensitive to magnetic field noise due to the surrounding nuclear spin bath and to charge noise.  Recent progress in this field has been discussed in Refs. \citet{2022Harvey,Laucht2021}.

\begin{figure*}[t!]
\resizebox{0.9\textwidth}{!}{\includegraphics{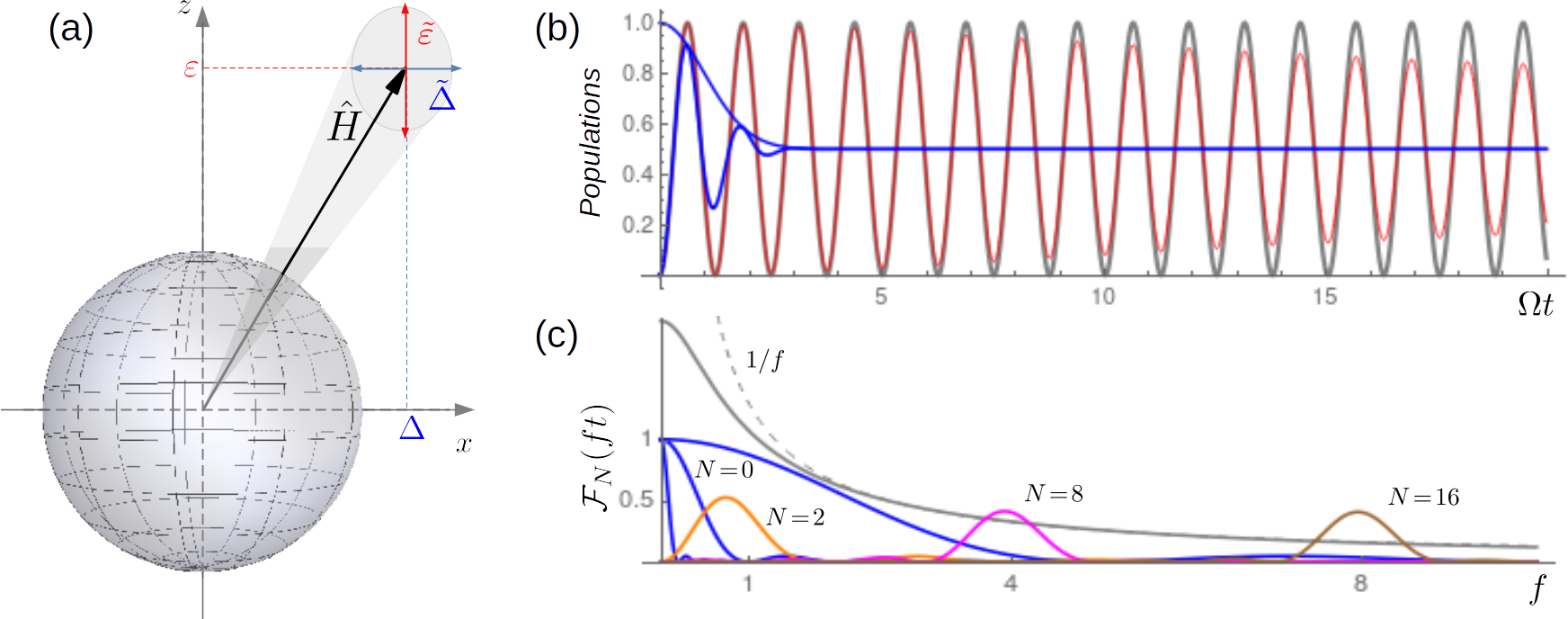}}
\caption{(a) In the spin language a pure state of a qubit is a unit vector of the Bloch sphere and the qubit Hamiltonian Eq.(\ref{eq:qubit}) is represented by a vector $\vec{H}$. Coherent oscillations correspond to a precession of the state vector about $\vec{H}$, whose direction and modulus fluctuate due to noise, affecting the qubit dynamics. (b) Coherent oscillations of the qubit populations in an isolated system (gray line) are strongly suppressed by longitudinal low-frequency noise (blue line), while the effect of transverse noise (red line) is much weaker. (c) Frequencies of the $1/f$ spectrum (grey lines, the thick one for a finite low-frequency cutoff) entering dephasing are filtered by the qubit dynamics: the blue curves represent ${\cal F}_0(f)$ entering Eq.(\ref{eq:gamma-long}) at different times, larger $t$ yielding smaller bandwidths.  
The orange, magenta and green curves represent the action of the filter Eq.(\ref{eq:filterfunc}) for periodic dynamical decoupling operated by $N$ $\pi$-pulses, $N=2$ representing spin-echo; the bandwidth migrates to larger frequencies suppressing the effect of low-frequency noise.  
\label{fig:coherentosc}}
\end{figure*}

\subsection{Quantum Sensing} \label{QS}
Coherent nanosystems are at the core of investigation also in the recent field of quantum sensing.
For a recent review, we refer the reader to Ref. \citet{Degen2017} and references therein. 
The idea behind is to turn into an exploitable property the sensitivity to external disturbances representing an intrinsic weakness for quantum computing purposes. Quantized energy levels, quantum coherent evolution or quantum entanglement are used to measure a physical quantity or to improve sensitivity beyond classical measurements.
To function as quantum sensors, physical systems need to satisfy the DiVincenzo criteria of efficient initialization and coherent manipulation \citep{DiVincenzo2000}. In addition, they have to interact with the physical quantity to be detected $V(t)$ which leads to  shifts of the quantized energy levels or splittings, $E(V)$.
The relevant quantity is the transduction parameter 
$\gamma = \partial^qE/\partial V^q$, where usually $q=1$ or $q=2$. In any physical implementation, the sensitivity to external signals carries also coupling with unwanted external noise sources. The trade-off between strong coupling to the quantity to be detected and weak disruptive noise effects are quantified by the sensitivity which scales as $\propto 1/ (\gamma \sqrt{T})$, where $T$ is the relevant decoherence or relaxation time \citep{Degen2017}. Quantum sensors require maximal transduction and long decoherence time. In general, the role of $1/f$ noise in these schemes is that of limiting the coherence time which governs the measurement time available for quantum sensing. Thus, lowering the amount of $1/f$ noise will increase the sensitivity and consequently also the dynamic range of quantum sensing (the largest signal is set by the smallest measurement time.)

The effect of low-frequency fluctuations can be illustrated in nitrogen-vacancy (NV) centers in diamond \citep{2013Doherty}, one of the most advanced platforms for quantum sensing. The use of electron spin states of NV-centers in diamond for highly sensitive magnetometry is an established fact with relevant applications. For an in depth description of this topic we refer to References  \citet{Rondin2014,Schirhagl2014,2020Rembold}. 
Key properties of these colour centers are the possibility to optically initialize, read out and coherently manipulate at ambient conditions.  The sensing protocol is essentially the measurement of the Zeeman splitting induced by an external magnetic field.
For advanced sensing applications, the transition frequency is detected via interferometric techniques. The basis of these techniques is Ramsey interference, the spin equivalent of an optical Mach–Zehnder interferometer.
The qubit is prepared in the superposition 
$|0 \rangle + e^{i \phi_0} |1 \rangle$ and Ramsey interference is based on  the phase accumulated by the Larmor precession during the interrogation time, This phase is related to the addressed transition frequency, which depends on the external field to be measured. This allows the detection of static,
slowly varying, or broadband near-DC signals. The main limitation of DC magnetometry with
NV centers is dephasing due to slowly varying inhomogeneities of the dipolar fields due to unpolarized spin impurities in the diamond crystal, mostly $^{13}$C and $^{14}$N. 

Dynamical decoupling (DD) schemes inspired by NMR \citep{kb:200-becker-highresNMR}, have been used to reduce the effect of low-frequency noise components \citep{Paladino2014}. These techniques are by now routinely employed both as active control tools for increasing the quality factors of quantum gates and in quantum sensors. The simplest instance is Hanh's echo, consisting of a $\pi$-pulse applied halfway of the detection in such a way as to reverse the qubit evolution so that the phases accumulated in the two-time intervals cancel out. 

DD refers to sequences of multiple $\pi$-pulses which allow shaping of the frequency response of the quantum sensor, in particular
attaining higher sensitivity when measuring AC magnetic fields and longer coherence times for quantum bits and gates. 

The DD control protocol is conveniently formulated in terms of filter functions entering the qubit coherent oscillations, analogously to Eqs. 
\ref{eq:gamma-long}, \ref{eq:gamma-transv}. For Gaussian noise coupled longitudinally, the filter function takes the form \citep{uhrig2007PRL}
\begin{equation}
\label{eq:filterfunc}
\begin{aligned}
\mathcal{F}_{N}(f,t)= \frac{1}{(2 \pi f t)^2} \big|1&+(-1)^{N+1} e^{i 2 \pi ft} 
\\&+ 2 \sum_{j=1}^N (-1)^j 
e^{i 2 \pi t_j}\big|^2  .
\end{aligned}
\end{equation}
$N$ is the number of pulses applied at time $t_j$ within the measurement time, $t$, perpendicularly with respect to the qubit quantization axis. A variety of pulse sequences have been investigated, differentiated by pulse directions and timings. The resulting filters effectively eliminate from the qubit evolution spectral noise components to different degrees (see Fig.~\ref{fig:coherentosc}c). For instance, the Carr–Purcell sequence \citep{carr1954PhysRev}, which consists of $n$ equidistant $\pi$ pulses, acts as a narrow quasi-monochromatic and tunable filter. The first pulse time, $t_1$ selects the pass-band frequency, while $n$ determines the
filter width. The CP sequences allow the detection of monochromatic AC fields with periodicity commensurate with the $2 t_1$, the interpulse timing. 
In NV sensor settings, periodic protocols have been exploited to detect and characterize individual $^{13}$C nuclear spins  placed a few nanometers apart from the NV center \citep{Zhao2012,Taminiau2012}.  

Despite the extraordinary progress, the sensitivities currently achieved are lower than
the theoretical limits established by the qubits' lifetimes. Material synthesis techniques have 
been developed to reduce paramagnetic impurities and strain effects in the host crystal.
In addition, quantum optimal control (QOC) methods have been applied to quantum sensors \citep{2011Giovannetti,2020Rembold,2021Hernandez}.
These techniques represent an alternative route to enhance the coherence time when noise at high-frequencies dominates.

\vspace{6pt}
\begin{center}
\emph{\small{Noise Spectroscopy}}
\end{center}
\vspace{2pt}

Noise spectroscopy is an important tool in quantum sensing. The term refers to methods for reconstructing the frequency spectrum of stochastic signals. The most common tools are based on Ramsey interferometry or more involved rf pulse sequences \citep{Degen2017}.
Noise spectroscopy can provide relevant information both on the signal to be detected and  on the intrinsic noise of the sensor. It can be formulated again in
terms of filter functions. In fact, in addition to the decoupling regime, 
there exist spectral regions of the filters of positive gain. The effect of the corresponding spectral components of noise is amplified \citep{Paladino2014}. 

Noise spectroscopy is routinely used as a powerful tool to characterize $1/f$ charge and magnetic flux noise in superconducting qubits. Here we only mention some recent results in this platform.
Exploiting the narrow-band filtering properties of the Carr Purcell Meibum Gill (CPMG) sequence \citep{meiboom1958RevSci}, authors of Ref. \citet{bylander2011NatPhys} measured flux noise in a persistent current qubit away from the optimal point in the spectral region $0.2–20$ MHz, where $1/f^{0.9}$ noise was detected. The decay of Rabi oscillations provided an independent measurement of the noise spectrum at frequencies  of the order of $\Omega_{Rabi} \approx$~MHz. Remarkably, the same power law behavior has been detected by Ramsey interferometry in the much lower frequency range $0.01 - 100$~Hz  in Ref.\citet{yan2012PRB}. By studying the decay of Rabi oscillations under strong driving  conditions
authors of Ref. \citet{Yoshihara2014} inferred flux noise up to $1.7$~GHz. Around $300$~MHz a spectrum close to
$1/f$ was detected where evidence of a Lorentzian component, possibly due to a more strongly coupled fluctuator was pointed out, see Figure 2.
The result of these types of measurements is sketched in Figure 1. 
Most noise spectroscopy protocols have been developed tailored to
Gaussian noise models. Norris and Viola in Ref. \citep{Norris2016} introduced a noise spectroscopy protocol
allowing for higher-order spectral  estimation of non-Gaussian noise, as demonstrated in a flux-tunable superconducting qubit sensor \citep{Sung2019}.
This represents a valuable tool for characterizing discrete microscopic sources with a $1/f$ spectrum.
More recently,  multilevel noise spectroscopy has been realized via a spin-locking-based protocol \citep{Sung2021}. 
The additional information gained from probing the higher-excited
levels enabled the identification of contributions from different underlying noise mechanisms in a flux-tunable transmon qubit. 
Noise spectroscopy tracking multilevel fluctuations represents also an efficient tool for materials characterization, as it has been demonstrated in a transmon fabricated with tantalum metal \citep{Tennant2022}.

Silicon quantum dot spin qubits have also been employed as  metrological devices to study the noise environment experienced by the qubit.
Noise spectroscopy has been done for an implanted phosphorus donor qubit in silicon (Si:P) \citep{2014Muhonen} and a SiGe quantum dot spin qubit  \citep{2018Yoneda}. 
Charge noise spectroscopy in an isotopically purified electron spin qubit in a $^{28}$Si/SiGe quantum dot
has been realized by an independent estimate of $S(f)$ around $0.01–1$ Hz by Ramsey interferometry, and around tens of kilohertz by CPMG sequence. Remarkably, the two estimates are consistent with $1/f^{1.01}$ spectrum extending over seven decades of frequencies \citep{2018Yoneda}. In this experiment, rapid spin rotations are applied by using a proximal micromagnet, which induces spin-electric coupling fields. In Ref. \citet{2018Chan} the environment of a  silicon metal–oxide–semiconductor (SiMOS) quantum dot spin qubit has been characterized via CPMG dynamical decoupling pulse sequences. 
The detected charge noise is $1/f^\alpha$ with $\alpha \approx 0.8 - 1$ for frequencies between $2$ and $20$ kHz.  Noise in the energy splitting of  an electron spin qubit in a  highly isotopically puriﬁed $^{28}$Si/SiGe device has been detected via Ramsey spectroscopy down to $10^{-5}$~Hz and by Hanh-echo at higher frequencies in Ref. \citet{Struck2020}. Measurements revealed a $1/f^2$ behavior below $5 \times 10^{-3}$ Hz followed by $1/f$ trend. These dependencies have been found  to be consistent
over 8 decades with charge noise independently detected in the same setup via measurement the current noise of a nearby SET sensor.

\section{Conclusion}

It is widely recognized that the road to avoiding $1/f$ noise in nanodevices is to improve the characteristics of the employed materials. Silicon crystal growth technology has been perfected in the past, and nowadays extremely pure silicon single crystals are available. Similar quality of materials would be required for various substances that are employed in quantum technology~\citep{deLeon2021}. One possible route for achieving high-quality materials is to tailor devices atom by atom \citep{ Khajetoorians2019}. The era of such a “designer matter” approach has recently begun, but plenty of work lies ahead before the quality and throughput of these materials will become relevant for alleviating $1/f$ noise in mainstream quantum science and technology. 

Progress in the more regular “engineered matter” is being made using composite, hybrid materials, for example, encapsulated  graphene with superconducting contacts \citep{Wang2019}. Such hybrid junctions have been employed in superconducting qubits, but it appears that there is still substantial $1/f$ noise present at the interfaces \citep{Haque2021}. In addition, different mechanisms may originate critical current fluctuations, for instance, charge carrier density fluctuations might also play a role \citep{Pellegrino2020}. The resistance noise properties of regular AlO$_x$ tunnel junctions are still superior, and the influence of their $E_J$ (or critical current, $I_c$) noise on qubit coherence time is small.  It will become essential when the coherence time will be in the ms range \citep{Siddiqi2021}. 

One of the emerging, striking fields of 2D materials is the magic angle twisted bilayer graphene (MATBG), in which superconductivity \citep{Cao2018} is assigned to flat energy bands \citep{Kerelsky2019,Xie2019,Torma2022}. The MATBG samples can be tuned between superconductivity and Mott insulating states with a rather small change in the charge carrier density. However, owing to variations in local charge doping or in twist angle, the superconducting properties of these samples may fluctuate in space or time, and thus display $1/f$ noise in critical current. A study of such noise will establish the quality of this material for certain sensitive detector applications, for example, their use in kinetic inductance detectors for radiation.

2D materials offer a platform for the miniaturization of components for coherent quantum circuits \citep{Antony2021,Wang_2022}. Large, low-loss capacitance per unit area can be achieved using a few layer hBN as an insulating layer. Furthermore, by utilizing h-BN as a gate dielectric, strong gate coupling for FETs can be obtained. Compared with amorphous hafnium oxide, the TLS density of h-BN dielectric is definitely smaller, which promises smaller $1/f$ noise for components in such 2D material technology.

With the increasing miniaturization of electronic quantum components, the relative magnitude of $1/f$ noise has been constantly growing over the years as the active part of the component is influenced strongly by one single impurity. Similar extreme purity as for single silicon crystals would be needed for other technologically relevant materials in the future \citep{deLeon2021}. An alternative solution to the $1/f$ noise dilemma is to harness independent single atoms or molecules \citep{Yu2021} as building blocks for quantum components and sensors. This has already been achieved quite well in quantum sensors based on NV centers in diamond or in optically read magnetometers based on rubidium gas for example \citep{Aslam2023}.

Besides dynamical decoupling schemes, the use of limited quantum resources can be bolstered by using phase estimation algorithms \citep{Kitaev1995}. These algorithms provide a sensitivity improvement directly proportional to the measurement time, i.e. they allow to achieve the Heisenberg limit in measurement accuracy. Combined with Bayesian learning algorithms, phase estimation schemes may provide ultimate speed for high-sensitivity sensing, e.g. magnetometry \citep{Bonato2016}, in spite of the influence of $1/f$ noise embedded in the coherence time of the sensing qubit. 

Quantum Optimal Control is another promising  methodology both for sensing applications and for quantum computation
applications which require high-fidelity gate operation. One instance where QOC can provide a solution is in reducing the cumulative errors due to hard pulses (i.e. rectangular pulses)  employed with DD sequences.

Fluctuations can provide important additional information in sensing, for example in gas recognition. Low-frequency noise due to adsorbant molecules and atoms on graphene has been demonstrated to yield distinct $1/f$ spectra  \citep{Rumyantsev2012}. In combination with supervised-learning algorithms, based on statistical learning theory, fluctuation-enhanced gas recognition has already been investigated \citep{Lentka2015}. Such an approach may provide a significant boost in gas recognition resolution for e-nose applications \citep{Chen2022}.

For future superconducting quantum technology, clean materials are essential as they increase coherence time, which improves qubit operation, and along with it, increases the sensitivity of all schemes for quantum sensing. One of the central issues in this quest for better superconducting junctions and conductors is to understand the microscopic origin of $1/f$ noise on interfaces, surfaces, and tunneling barrier dielectrics. This will call for the systematic development of materials growth, cleaning, and fabrication techniques, as well as systematic benchmarking of the realized quantum components.

\bibliography{bib-oneoverf}

\end{document}